\documentclass[twocolumn,prl,superscriptaddress,nofootinbib,preprintnumbers]{revtex4}
\usepackage{graphicx}
\usepackage{color}
\usepackage{amsmath} 
\usepackage{amssymb} 
\usepackage{bm}
\usepackage{slashed}
\usepackage{hyperref}
\usepackage{ifpdf}
\usepackage{latexsym}
\usepackage{bm}

\definecolor{nicered}{rgb}{0.7,0.1,0.1}
\definecolor{nicegreen}{rgb}{0.1,0.5,0.1}
\hypersetup{colorlinks,citecolor= nicegreen,linkcolor= nicered}

\newcommand{\cut}[1]{{}}

\def\beq{\begin{equation}}
\def\eeq{\end{equation}}
\newcommand{\ba}{\begin{array}}
\newcommand{\ea}{\end{array}}
\newcommand{\bea}{\begin{eqnarray}}
\newcommand{\eea}{\end{eqnarray} }
\newcommand{\bal}{\begin{align}}
\newcommand{\eal}{\end{align}}
\def\bi{\begin{itemize}}
\def\ei{\end{itemize}}
\def\ben{\begin{enumerate}}
\def\een{\end{enumerate}}
\def\beq{\begin{equation}}
\def\eeq{\end{equation}}
\def\bc{\begin{center}}
\def\ec{\end{center}}
\def\bt{\begin{table}}
\def\et{\end{table}}
\def\btb{\begin{tabular}}
\def\etb{\end{tabular}}

\def\mass2{mass${}^2$}

\def\mass2{mass${}^2$}

\def\im{{\rm Im}\,}

\def\simlt{\stackrel{<}{{}_\sim}}
\def\simgt{\stackrel{>}{{}_\sim}}

\def\eps{\epsilon}

\begin{document}
\preprint{CERN-PH-TH/170}
\preprint{DESY 11-110}

\title{Constraining the dipole moments of the top quark}%

\author{Jernej F. Kamenik}
\email[Electronic address:]{jernej.kamenik@ijs.si} 
\affiliation{J. Stefan Institute, Jamova 39, P. O. Box 3000, 1001 Ljubljana, Slovenia}
\affiliation{Department of Physics,
  University of Ljubljana, Jadranska 19, 1000 Ljubljana, Slovenia}

\author{Michele Papucci}
\email[Electronic address:]{mpapucci@lbl.gov} 
\affiliation{Ernest Orlando Lawrence Berkeley National Laboratory, University of California, Berkeley, CA 94720}
\affiliation{CERN, Theory Division, CH-1211, Geneva 23, Switzerland}
\author{Andreas Weiler}
\email[Electronic address:]{andreas.weiler@desy.de} 
\affiliation{CERN, Theory Division, CH-1211, Geneva 23, Switzerland}
\affiliation{DESY, Notkestrasse 85, D-22607 Hamburg, Germany}
\date{\today}

\begin{abstract}
We investigate the direct and indirect bounds on dipole operators involving the top quark. A careful analysis shows that the experimental upper limit on the neutron electric dipole moment strongly constrains the chromo-electric dipole of the top. We improve previous bounds by two orders of magnitude. This has significant implications for new physics models and it also means that CP violation in top pair production mediated by dipole operators will not be accessible at the LHC.
The CP conserving chromo-magnetic dipole moments are constrained by recent measurements of the $t\bar t$ spectrum by the ATLAS collaboration. We also update the indirect constraints on  electric and magnetic dipole moments from radiative $b\rightarrow s$ transitions, finding that they can be  considerably larger than their colored counterparts.
\end{abstract}

\maketitle

\def\simgt{\mathrel{\lower2.5pt\vbox{\lineskip=0pt\baselineskip=0pt
         \hbox{$>$}\hbox{$\sim$}}}}
\def\simlt{\mathrel{\lower2.5pt\vbox{\lineskip=0pt\baselineskip=0pt
         \hbox{$<$}\hbox{$\sim$}}}}

\newcommand{\mA}{m_{A^0}}
\newcommand{\Tr}{\rm Tr}
\newcommand{\luv}{\Lambda_{UV}}
\newcommand{\Np}{N^\prime}
\newcommand{\Lh}{\Lambda_H}
\newcommand{\Ls}{\Lambda_S}
\newcommand{\Lg}{\Lambda_G}
\newcommand{\cuFF}{{\cal C}_{m_u}}
\newcommand{\cdFF}{{\cal C}_{m_d}}
\newcommand{\cudF}{{\cal C}_{\mu}}
\newcommand{\cudFb}{\bar{\cal C}_{\mu}}
\newcommand{\cudFF}{{\cal C}_{B_\mu}}

\newcommand{\be}{\begin{equation}}
\newcommand{\ee}{\end{equation}}

\subsection*{Introduction}
In many scenarios the top quark provides a preferred window on physics beyond the Standard Model (SM), given its large coupling to the physics responsible for the Electro-Weak Symmetry Breaking (EWSB). Top quarks will be copiously produced at the LHC -- the average rate in September 2011 - during which the nominal luminosity at 7 TeV has been reached -  has been of the order of 30 tops per minute, and is expected to reach 9 tops per second at design luminosity $10^{34}~\rm cm^{-2} s^{-1}$ and 14~TeV center of mass energy. While the top quark has been studied in some detail at the Tevatron, many of its properties besides the mass, spin and the color and electric charges are still poorly constrained. Significant new insights on top quark properties will therefore be one of the tasks of the LHC. 

A fascinating possibility is that the top quark shows deviations from its behavior predicted by the SM.
The  leading  contributions are encoded in the (chromo)-electric and (chromo)-magnetic dipole moments, (C)EDM and (C)MDM in the following. 

A particularly interesting scenario is realized if some of the quarks are partially or
fully composite~\cite{Kaplan:1991dc}. The top, being the most massive quark, is the most
natural candidate, as the mass and the amount of compositeness are often related, see e.g.~\cite{rscomp}. 
If the compositeness scale is in its natural range, large
CMDM and CEDM are expected~\cite{rscedm}. 

Supersymmetric models on the other hand, can also lead to enhanced dipole moments of the top. This happens if 
the supersymmetric partners of the top are not too heavy, which is a requirement of naturalness and possibly electro-weak baryogenesis, see e.g.~\cite{Carena:2008vj}. The presence of sizable flavor-blind phases accessible to the third-generation quarks can also explain the recent hints of CP violation in $B_{s}$ mixing in the context of Minimal Flavor Violation~\cite{Kagan:2009bn}.

We leave the implications of our results for specific scenarios to future work.
In this paper we investigate the direct and indirect constraints on dipole operators involving the top quark in a model independent way. In particular we will critically reanalyze the phenomenologically relevant question whether CP violation in top pair production can be mediated by dipole operators at a level accessible for the LHC. 

As it is well known, dipole moments before EWSB are encoded in dimension-6 operators 
\begin{align}
&\left(c_{LR,c} \,\, g_{s}\bar Q H\,\sigma^{\mu\nu}T^{a}U +{\rm h.c.}\right) G_{\mu\nu}^{a}\,,  \nonumber \\
& \left(c_{LR,w}\,\, g \,\bar Q \,\tau^{a}H\sigma^{\mu\nu}U +{\rm h.c.}\right) W_{\mu\nu}^{a}\,, \nonumber \\
&  \left(c_{LR,y}\,\, g' \bar Q H\,\sigma^{\mu\nu}U +{\rm h.c.} \right) B_{\mu\nu}\,,
\label{eq:Lgauge}
\end{align}  
with dimensional couplings $c_{i}\sim 1/\Lambda_{i}^{2}$. In particular, $\im(c_{LR,i})\neq0$ would signal CP violation.

In the following, we will employ a phenomenological Hamiltonian which can be easily translated to the more physical $SU(2)\times U(1)$ gauge invariant basis in Eq.~(\ref{eq:Lgauge}), 
\begin{align}
\mathcal H_{\rm eff} =
- \frac{1}{2} 
\bar \psi_q \,\Big[& (F_{\mu\nu} \sigma^{\mu\nu}) (\mu_q + i \gamma_5 d_q)
\nonumber\\
+ g_s &\left.   (G_{\mu\nu}^a t^a \sigma^{\mu\nu}) (\tilde \mu_q + i \gamma_5 \tilde d_q) \right]\,\psi_q \nonumber\\
& 
- \frac{1}{6} w f^{abc} \varepsilon^{\mu\nu\lambda\rho}\, G_{\mu\sigma}^a G_{\nu}^{b\sigma} G_{\lambda\rho}^c\,,
\label{eq:Leff}
\end{align}
where $q=u,d,s,c,b,t$ and $\varepsilon^{0123}=1$. We denote $d_q$ and $\tilde d_q$ as the EDM and the CEDM of the quark $q$, while $\mu_q$ and $\tilde \mu_q$ are the corresponding MDM and CMDM. We have included the CP violating Weinberg operator~\cite{Weinberg:1989dx} (contributing with the Wilson coefficient $w$), which will be crucial later. Correspondingly, we have omitted the terms involving the charged gauge bosons, since they will not play an important role in the following and have been already investigated elsewhere~\cite{tbw}. In the following analysis we will consider the contributions of a single operator at a time. Consequently, the derived constraints will apply in absence of cancellations among several conspiring operator contributions.

\subsection*{Indirect constraints}

\begin{figure}
\includegraphics[scale=1]{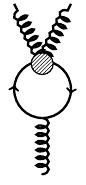}\qquad
\includegraphics[scale=1]{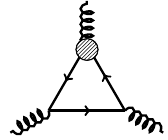}
\caption{ Diagrams generating the contribution to the Weinberg operator at the top threshold. The grey blob denotes the insertion of the chromo-electric dipole operator~\cite{Braaten:1990gq,Chang:1991ry,Boyd}\label{fig:threshold}.}
\end{figure}

We first consider the present indirect constraints on the top CEDM $\tilde d_t$.
The operators in Eq.~(\ref{eq:Leff}) run and mix under QCD renormalization group (RG) evolution. At present, these effects are known to NLL accuracy~\cite{Degrassi:2005zd}. In particular the Weinberg operator mixes into the (C)EDMs of quarks, but not vice versa. Nevertheless, it has been known for some time~\cite{Braaten:1990gq,Chang:1991ry} that the CEDMs induce a \emph{finite} threshold correction to the Weinberg operator when a heavy quark is integrated out, as shown in Fig.~\ref{fig:threshold},
\be
\delta w^{(q)} = \frac{g_s^3}{32\pi^2} \frac{\tilde d_q}{m_q}\,,
\ee
where $g_s$ and $m_q$ are evaluated at the heavy quark threshold scale.
In the case of the top quark, the combined effects of the finite shift in $w$ and the subsequent RG evolution to the hadronic scale will induce non-zero contributions also for the (C)EDMs of the light quarks. 

Performing the QCD evolution of the Weinberg operator at NLL accuracy down to the hadronic scale $\mu_H\sim 1$~GeV, taking into account the relevant $m_b$ and $m_c$ thresholds, we thus obtain $d_{u,d}(\mu_H)$, $\tilde d_{u,d}(\mu_H)$ and $w(\mu_H)$ in terms of $\tilde d_t(m_t)$ 
\begin{align}
d_u &= -3.1 \cdot 10^{-9}\,e \, \tilde d_t\,, & d_d &= 3.5 \cdot 10^{-9}\,e \, \tilde d_t \,, \nonumber\\
\tilde d_u &= 8.9 \cdot 10^{-9}\, \tilde d_t \,,& \tilde d_d &= 2.0 \cdot 10^{-8} \, \tilde d_t\,, \nonumber\\
w &= 1.0 \cdot 10^{-5}~{\rm GeV}^{-1} \, \tilde d_t\,,
\label{eq:dx}
\end{align}
where we have used $m_t=173.3$~GeV~\cite{:1900yx}, $m^{\overline{MS}}_d(2~{\rm GeV}) = (4.7\pm 0.1)$~MeV, $m^{\overline{MS}}_u(2~{\rm GeV})=(2.1\pm 0.1)$~MeV~\cite{Laiho:2011rb} and $\alpha^{\overline{MS}}_s(m_Z)=0.1184$~\cite{Bethke:2009jm}.

Presently, the most sensitive observables are the atomic EDMs of mercury ($d_{Hg}<3.1\cdot 10^{-29}$~e~cm\, at $90\%$\,C.L.~\cite{Griffith:2009zz}) and of the neutron ($d_n < 2.9\cdot 10^{-26}$~e~cm\, at $90\%$\,C.L.~\cite{Baker:2006ts}). Following~\cite{Pospelov:2005pr}, we  evaluate the relevant contributions as
\begin{subequations}
\begin{align}
d_{Hg} =& - 1.8 \cdot 10^{-4}\, {\rm GeV}^{-1} \, e\,  \bar g^{(1)}_{\pi N N}\,,\\
d_n =&  (1\pm 0.5 )\,[ 1.1 \,e\,( \tilde d_d + 0.5 \,\tilde d_u ) + 1.4 \,( d_d - 0.25 \,d_u) ]
 \nonumber\\
&+ (22\pm 10)\cdot 10^{-3}\, {\rm GeV} \, e\, w\,,
\end{align}
\end{subequations}
where $g^{(1)}_{\pi NN} = 4^{+8}_{-2} \,(\tilde d_u - \tilde d_d)\rm \, GeV$. All  quantities are evaluated at the scale $\mu_H\sim 1$~GeV. The values and uncertainty estimates for the relevant matrix elements, particularly for the Weinberg operator contribution to $d_n$, have been evaluated using QCD sum rule techniques~\cite{Demir:2002gg}.

Inserting (\ref{eq:dx}) into the above expressions and treating all the relevant theoretical uncertainties as flat distributions within the stated errors, we find that the neutron EDM constrains the top CEDM to be
\be
|\tilde d_t| < 2.1\cdot 10^{-19}\, \rm{cm}~~  \, (90\%\,\rm{C.L.})\,,
\label{eq:ind}
\ee
i.e. $|\tilde d_{t} \,m_{t}|<1.9\cdot 10^{-3}$. The constraint from the neutron EDM is dominated by the contribution to the Weinberg operator, which amounts to roughly $85\%$ of the total effect of $\tilde d_t$ in $d_n$, even though the light quark (C)EDM contributions are not totally negligible. Furthermore, the constraint from $d_{Hg}$ provides a (two orders of magnitude) weaker bound on $\tilde d_{t}$, since it is not sensitive to the Weinberg operator and also comparatively weaker than $d_{n}$ for the light quark CEDMs. We note in passing the a similar constraint on the EDM of the $b$ quark has previously been obtained~\cite{Chang:1991ry}.

\begin{figure}[!t]
\begin{center}
\includegraphics[scale=1.0]{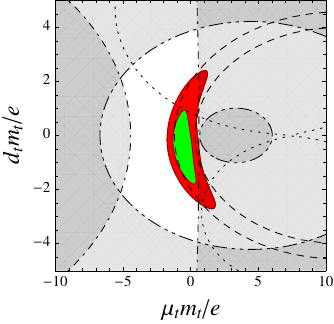}
\end{center}
\caption{\label{fig:1} Indirect constraints on the top EDM and MDM coming from radiative and rare semileptonic B decay observables (as defined in Ref.~\cite{Altmannshofer:2012az}): $Br(B\to X_s\gamma)$ (dashed $1\sigma$ contours), $A_{CP}(B\to X_s \gamma)$ (dotted $1\sigma$ contours), $\langle A_{FB}\rangle(B\to K^* \ell^+\ell^-)[1\,{\rm GeV}^2<q^2<6\,{\rm GeV}^2]$ (dot-dashed $1\sigma$ contours) and $\langle F_{L}\rangle(B\to K^* \ell^+\ell^-)[1\,{\rm GeV}^2<q^2<6\,{\rm GeV}^2]$ (double dot-dashed $1\sigma$ contours)\,. The combined allowed region is bounded by full line contours and shaded in lighter green ($68\%$C.L.) and darker red ($95\%$C.L.).}
\end{figure}

The indirect constraints on the other top dipole moments in Eq.~(\ref{eq:Leff}) are considerably weaker. The EDM of the top, $d_{t}$, induces  light quark EDMs only through weak interactions and is suppressed by flavor mixing factors~\cite{CorderoCid:2007uc} resulting in $d_d = 2.4 \times 10^{-12} \,d_t$\,, and consequently we find a weak bound of 
\be
|d_{t}|<1.7 \times 10^{-14}\, \rm e\,cm~~ \, (90\%\,\rm{C.L.})\,.
\ee
A stronger limit comes from $b\rightarrow s\gamma$ and $b\rightarrow s \ell^{+}\ell^{-}$ processes, since the leading SM contribution carries the same loop and flavor suppressions. Following~\cite{Hewett:1993em}, we obtain
\begin{align}
& \Delta C_{7\gamma}(m_{W})=6.5 \cdot 10^{-2}\left( \mu_{t} - 2.65 i \, d_{t}\right)m_{t}/e\,,
\end{align}
where we have included the effects of the top MDM which is also constrained. $\Delta C_{7\gamma}$ is the new contribution to the Wilson coefficient of the magnetic operator mediating the $b\rightarrow s $ transition. 
Using the results of a global fit to radiative and rare semileptonic $B$ decays~\cite{Altmannshofer:2012az}, we obtain the allowed region in the $(\mu_t, d_t)$ plane in Fig.~\ref{fig:1}. 
The most sensitive observables are $Br(B\to X_s \gamma)$, $\langle A_{FB}\rangle(B\to K^* \ell^+\ell^-)[1\,{\rm GeV}^2<q^2<6\,{\rm GeV}^2]$ and $\langle F_{L}\rangle(B\to K^* \ell^+\ell^-)[1\,{\rm GeV}^2<q^2<6\,{\rm GeV}^2]$ (all defined in Ref.~\cite{Altmannshofer:2012az}). Marginalizing over one of the moments, we can obtain a two-sided bound for the other, leading to 
\begin{subequations}
\begin{align}
-1.5&< \mu_{t}m_{t}/e <1.0~( 95\%\,{\rm C.L.}),\\
-2.3&< d_{t}m_{t}/e <1.7~( 95\%\,{\rm C.L.})\,.
\end{align}
\end{subequations}
In the future, improved precision on the experimental side (as expected from Super B Factories~\cite{superB}) together with theoretical refinements could lead to a slightly more stingent bound.
These processes are also sensitive to the top CMDM, through the operator ${\cal O}_{8g}$. However, due to the additional $\alpha_{s}$ suppression, the indirect bound is significantly weaker and will be superseded by the direct one derived in the next section.

\subsection*{Collider constraints}

The gluonic dipole couplings of the top directly affect $t\bar t$ production at hadron colliders~\cite{colliderdipole}.
The complementarity of the total production cross-section measurements at the Tevatron and the LHC in constraining such contributions has recently been pointed out in~\cite{Hioki:2009hm}. In~\cite{Degrande:2010kt}, the CMDM was constrained using Tevatron data in combination with four fermi operators and spin correlations were shown to be promising signatures at the LHC. In addition, a number of $T$-odd correlation observables can be constructed which are sensitive to the top CEDM  (c.f.~\cite{cptopreview}). 
We comment on the future prospects of these observables at the end of the paper. 

At present, the most sensitive observables available at the Tevatron and the LHC are the total production cross-sections and their differential spectrum as a function of the invariant mass of the top pair ($m_{t\bar t}$).

At the Tevatron the recent combination of the CDF analyses yields~\cite{ttxsectionTevatron},
\begin{align}
\sigma^{\rm Tevatron}_{\rm exp} = (7.50\pm 0.48){\rm~pb} \,,
\end{align}
for an assumed top mass of $m_t=172.5$~GeV. We have combined the estimated statistical and systematic errors in quadrature. This is consistent with the most recent theoretical SM prediction
for this observable~\cite{nnlottbar},
\begin{align}
\sigma^{\rm Tevatron}_{\rm SM} = (6.75^{+0.08}_{-0.42}){\rm~pb}  \,,
\end{align}
based on approximate NNLO QCD calculation using the same top mass and the MSTW2008 PDFs~\cite{Martin:2009iq}\footnote{Different approximations to the full NNLO QCD result employed by the present calculations give mildly inconsistent results for $\sigma^{\rm Tevatron}_{\rm SM}$, while the agreement at high $m_{t\bar t}$ as well as for the LHC observables is much better (see~\cite{Kidonakis:2011ca} for a recent review). We have checked explicitly, that our bounds are not significantly affected by using different approximate NNLO QCD results~\cite{Kidonakis:2011ca}.
}. As pointed out in~\cite{Blum:2011up}, the most significant information in the high $m_{t\bar t}$ region at the Tevatron is the one derived from the next-to-highest measured bin~\cite{Aaltonen:2009iz},
\begin{align}
\sigma{(700~{\rm GeV}<m_{t\bar t}<800~{\rm GeV})}_{\rm exp} = (80\pm37)~\rm fb\,, 
\end{align}
to be compared with the SM theory prediction of~\cite{Ahrens:2010zv},
\begin{align}
\sigma(700~{\rm GeV}<m_{t\bar t}<800~{\rm GeV})_{\rm SM}=(80 \pm 8\rm)~fb \,. 
\end{align}

\begin{center}
\begin{figure}
\includegraphics[scale=1.25]{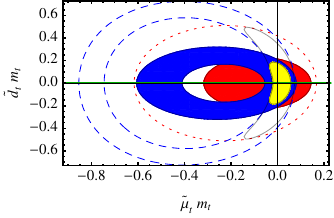}
~\\
~\\
\includegraphics[scale=1.26]{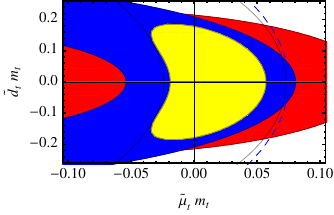}
\caption{
Combined current LHC and Tevatron $95\%$ C.L. constraints on the top CMDM ($\tilde \mu_t$) and the CEDM ($\tilde d_t$)  (shaded in yellow). Individual constraints come from the total cross-section and $m_{t\bar t}$ spectrum measurements at the Tevatron (dashed blue and doted red), as well as the LHC (shaded blue and red). The combination of only Tevatron constraints is drawn in black. Finally the CEDM indirect constraint is presented in green.
\label{fig:collider}}
\end{figure}
\end{center}

At the LHC at 7 TeV, presently the most precise measurement~\cite{ttATLAS} for the total production cross-section yields
\begin{align}
\sigma_{\rm exp}^{\rm LHC} = ( 180\pm18 )\rm~pb, 
\end{align}
 in agreement with the SM prediction of~\cite{Aliev:2010zk},
\begin{align} 
 \sigma_{\rm SM}^{\rm LHC} = (165^{+11}_{-16})\rm~pb. 
\end{align} 
 Recently the ATLAS collaboration published a study~\cite{ATLAS-CONF-2011-087} of the $m_{t\bar t}$ spectrum using $200$~pb${}^{-1}$ collected luminosity. In the high $m_{t\bar t}>1$~TeV region they report $N_{\rm exp}^{1\rm TeV} = 77\pm 9$ events. The results were not unfolded. 
Among the systematic uncertainties, the $b$-tagging efficiency ($11\%$) dominates the total inclusive cross-section measurement and stays almost constant with $m_{t\bar t}$.  The other important source of error is the jet energy scale uncertainty. It is subleading in the total inclusive measurement ($9\%$), with a mild $m_{t\bar t}$ dependence. 
Since the ATLAS result is not unfolded, one needs to take into account the invariant mass resolution, and the reconstruction efficiency and acceptance, $(A \cdot \epsilon)$. We model the former by smearing our partonic $m_{t\bar t}$ distributions with a Gaussian kernel. We estimate a $\mathcal O(0.25 \cdot m_{t\bar t})$ width for this smearing by comparing to the reconstructed invariant mass resolutions of a sample of narrow $Z'$ models in the same ATLAS study. Finally, to estimate $A\cdot\epsilon$ we compare the smeared SM $m_{t\bar t}$ distribution, computed using known approximate NNLO QCD results~\cite{ Ahrens:2010zv}, with the reconstructed SM background distribution presented in the ATLAS Note. Using $Br(t\bar t \to 4j+\ell)=0.3$ (where $\ell=e,\mu$) we extract a constant $A\cdot \eps \simeq 0.3$ for the $m_{t\bar t}$ bins between $1~{\rm TeV}< m_{t\bar t} < 1.6~\rm TeV$, which can be now used to compare the signal with the data.

We find for the measured $t\bar t$ cross-sections 
including statistical and our estimates for the systematic uncertainties 
\begin{subequations}
\begin{align}
\sigma(1~{\rm TeV}<m_{t\bar t}<1.2~{\rm TeV}) &= (2.9\pm 0.6)~\rm{pb}\,, \\
\sigma(1.2~{\rm TeV}<m_{t\bar t}<1.4~{\rm TeV}) &= (1.0\pm 0.3)~\rm{pb}\,, \\
\sigma(1.4~{\rm TeV}<m_{t\bar t}<1.6~{\rm TeV}) &= (0.45\pm 0.19)~\rm{pb}\,,
\end{align}
\end{subequations}
which corresponds to
\beq
\sigma(m_{t\bar t}>1~{\rm TeV}) = (4.5\pm 0.79)~\rm pb\,.
\eeq
These results may be directly compared to partonic $m_{t\bar t}$ distributions smeared with  $0.25 \cdot m_{t\bar t}$-wide Gaussians.

We evaluate the effect of the top CMDM and CEDM on the relevant Tevatron and LHC observables at LO in QCD, using the known partonic cross-section formulae~\cite{colliderdipole} convolved with MSTW2008 PDFs~\cite{Martin:2009iq}\,. We normalize our SM values to respective approximate NNLO results~\cite{nnlottbar, Ahrens:2010zv, Aliev:2010zk} including theoretical uncertainties. 
Additionally we have checked the residual theoretical uncertainty in the relative NP contributions by varying the factorization and renormalization scales and finding negligible differences.
We compare these estimates of the inclusive and high $m_{t\bar t}$ cross-sections with the corresponding measurements both at the Tevatron and the LHC in Fig.~\ref{fig:collider}. 

We observe that the new ATLAS result on the high $m_{t\bar t}$ region at the LHC sizably shrinks the allowed region in the $(\tilde \mu_t, \tilde d_t)$ plane relative to previous results~\cite{Hioki:2009hm} or compared to using only Tevatron data. 
Marginalizing over the CEDM values, we obtain a new best bound on the top CMDM of
\be
|\tilde \mu_t | m_t < 0.05~~ \,  (95\%\,\rm{C.L.})\,.
\ee
The CEDM of the top is constrained to 
\be
|\tilde d_t | m_t < 0.16~~ \,  (95\%\,\rm{C.L.})\,,
\ee
or $|\tilde d_t| < 1.9 \cdot 10^{-17}$~cm, which is almost two orders of magnitude weaker than our new indirect bound (\ref{eq:ind}). A remark on the consistency of our EFT expansion is in order here. The CP violating CEDM does not
interfere with the SM and its contribution to the cross section starts at $\sim 1/\Lambda^4$, and not at $\sim 1/\Lambda^2$ as the CP conserving CMDM. It is therefore of the same order as interfering dimension eight operators, which in principle should have been included. Fortunately the far dominant constraint arises from  indirect observables and we can safely ignore this issue. 

The bounds can be expressed in terms of the gauge-invariant basis of Eq.~(\ref{eq:Lgauge}). The minimum scale for new physics contributing to the gluonic dipole moments has to be 
\begin{align}
{\rm Re}\, \Lambda_{LR,c}^{\rm direct}&> 1.1 \,\,{\rm TeV}\,,\nonumber\\
{\rm Im} \, \Lambda_{LR,c}^{\rm direct}&> 0.62\, \,{\rm TeV}\,,\nonumber\\
{\rm Im} \, \Lambda_{LR,c}^{\rm neutron}&> 4.7 \,\,{\rm TeV}\,,\nonumber
\end{align}
where we have separated the minimum scale for  real and imaginary contributions.

Let us now consider the prospects of probing a CEDM at the LHC as small as required by the indirect bound~(\ref{eq:ind}) in absence of large cancellations. In~\cite{Gupta:2009wu}, it was estimated that a $5\sigma$ detection for a value of $\tilde d_{t} m_{t} =0.05$ would require $10\,\rm fb^{-1}$ at $14$~TeV. We see that one would therefore need at  least ${\cal O}(1\,{\rm ab^{-1}})$ luminosity to detect CP violation at $5\sigma$ from a top CEDM. This is likely beyond  current LHC capabilities, unless further collider studies can improve the reach (c.f.~\cite{SLHC}).

\subsection*{Conclusions}
We have investigated the constraints on dipole operators involving the top quark. In absence of large cancellations, the neutron EDM imposes a stringent bound on the top CEDM. We have also derived a direct bound on the CMDM of the top from the invariant mass distribution in $t \bar t$ events at the LHC. These two bounds have significant implications for models of top compositeness and supersymmetric models with light stops which we will address in an upcoming paper. 
Finally, we have updated the constraints on the EDM/MDM of the top coming from $b\rightarrow s $ transitions with the most recent data.
\section*{Acknowledgements}
This work was initiated at the tmini workshop at the Weizmann Institute of Science. We thank the organizers for the inspiring atmosphere and great hospitality.
We thank Christophe Grojean and Roberto Contino for pointing out typos in Eqs.~(6) and (8) in the first version of the paper. We would also like to thank Georgios Choudalakis, Elin Bergeaas Kuutmann, Michele Redi,  Pekka Sinervo, David Straub and Joaquim Matias for useful discussions.  M.P. would like to thank the Aspen Center for Physics where part of this work was completed. 

The work of M.P. was supported in part by the Director, Office of Science, Office of High Energy and Nuclear Physics, of the US Department of Energy under Contract DE-AC02-05CH11231.  The work of J.F.K. was supported in part  by the Slovenian Research Agency. The work of A.W. was supported in part by the German Science Foundation (DFG) under the Collaborative Research Center (SFB) 676.


\end{document}